\documentclass[a4paper]{article}

\usepackage{INTERSPEECH2022}
\usepackage{bbding} 
\usepackage{cite} 

\title{Meta Auxiliary Learning for Low-resource Spoken Language Understanding}
\name{Yingying Gao, Junlan Feng, Chao Deng, Shilei Zhang}
\address{
  China Mobile Research Institute }
\email{\{gaoyingying, fengjunlan, dengchao, zhangshilei\}@chinamobile.com}

\begin{document}

\maketitle
\begin{abstract}
Spoken language understanding (SLU) treats automatic speech recognition (ASR) and natural language understanding (NLU) as a unified task and usually suffers from data scarcity. We exploit an ASR and NLU joint training method based on meta auxiliary learning to improve the performance of low-resource SLU task by only taking advantage of abundant manual transcriptions of speech data. One obvious advantage of such method is that it provides a flexible framework to implement a low-resource SLU training task without requiring access to any further semantic annotations. In particular, a NLU model is taken as label generation network to predict intent and slot tags from texts; a multi-task network trains ASR task and SLU task synchronously from speech; and the predictions of label generation network are delivered to the multi-task network as semantic targets. The efficiency of the proposed algorithm is demonstrated with experiments on the public CATSLU dataset, which produces more suitable ASR hypotheses for the downstream NLU task.
\end{abstract}

\section{Introduction}

Spoken language understanding (SLU) aims at inferring semantic information from spoken utterances. It is typically posed as two subtasks: utterance-level intent classification, and sequence tagging for slot types and values. 

A standard SLU system comprises two components: an automatic speech recognition (ASR) component to generate text or N-best hypotheses from speech and a natural language understanding (NLU) component to extract semantic constituents from the the ASR output. 
Each component is trained and optimized separately, so that the upstream ASR outcomes might degrade the performance of NLU and the desired NLU predictions would not inform ASR model to focus on these tokens. The recent end-to-end (E2E) approaches address this problem via predicting SLU output directly from speech inputs \cite{end1,end2,end3,end4,end5,end6,end7,end8,end9}. However, these methods require a large amount of semantically labelled speech data to exceed the cascaded ASR-NLU systems and the data collection is time-consuming and expensive. 

The first solution to address data sparseness problem is transfer learning \cite{transfer1,transfer2,transfer3,transfer4}. \cite{transfer1} proposes a simple end-to-end crosslingual spoken language understanding model based on a pretrained mutli-language model in 53 languages - XLSR53, which achieves state-of-the-art performance on the Fluent Speech Commands (FSC) English database and competitive results on the CATSLU-MAP Chinese database. \cite{transfer2} adopts Teacher-Student learning to align the SLU output space to the output space of a pre-trained NLU model so as to realize few-shot transfer. 
The second solution to prevent the end-to-end system from over-fitting is data augmentation with synthetic datasets \cite{aug1,aug2}. In \cite{aug1}, given a dialogue act, corresponding utterances are generated based on a pretrained general language model SC-GPT, in order to increase the corpus variability.
In \cite{aug2}, speech synthesis is used to generate synthetic training data and the effectiveness is confirmed on two open-source SLU datasets.
The third solution for the training of low resource SLU model is integrating knowledge as additional input or auxiliary task to learn \cite{phone1,phone2,history1,history2}.
\cite{history2} takes the decoded sequence and SLU labels of previous turns as dialog history, then encodes the history as BERT embeddings and utilizes it as an additional input along with speech features of the current utterance. 
\cite{phone1} proposes a joint textual-phonetic pre-training approach to learn spoken language representation, in the hope to explore the potentials of phonetic information on the improvement of SLU robustness to ASR errors. 

The aforementioned auxiliary-task or multi-task learning is considered as a compensation for the primary task which may be data hungry. However, it requires corresponding tags for the auxiliary task and the scarcity problem still exists. In this paper, we introduce Meta AuXiliary Learning (MAXL) \cite{maxl} to SLU network training, which automatically learns appropriate labels for the auxiliary task without requiring any further annotations. Specifically, two networks are trained and optimized jointly: a NLU network to predict intent and slot tags from texts, and a multi-task network to train ASR task along side SLU task. The NLU network and the ASR sub-task can be pre-trained via text-to-label data and speech-to-text data independently to pursuit a better performance. During joint training, the ground truth SLU tags are not required and the SLU targets in multi-task network are supported by the label generation network. 
The interaction between the two networks is produced by the SLU tags supplied via the label generation network. Therefore, in order to allow gradient flow in back propagation, we investigate different interfaces between the two networks and analyze their impacts on both ASR performance and NLU accuracy. The purpose of the proposal is to enhance the SLU prediction with limited SLU tags. The multi-task network makes the ASR task focus on the tokens related with SLU predictions. The incorporation of the losses from the label-generation network and the multi-task network lead to a double gradient during the updating of label-generation network, which makes it as a form of meta learning. Due to this interaction, the label-generation network absorbs knowledge from the ASR network and obtains better performance. Meanwhile, the ASR outputs are more suitable for the downstream NLU task. 

The major contribution of this paper is two-fold:

1) We introduce a meta-learning method to train ASR model and NLU model jointly without semantic tags on speech data, improving the quality of the ASR outcomes on downstream task and acquiring knowledge from the supervised task for NLU.

2) We study the interactions between NLU and ASR in a bidirectional way and investigate different interfaces from NLU network to ASR-SLU network to make the entire network differentiable and to realize the joint training of the two networks in a low-resource situation.

\section{Related work}

Meta learning is the recent popular method in low-resource settings \cite{meta1,meta2,meta3,meta4,meta5,meta6,meta7}. \cite{meta1} adopts meta learning on a code-switching speech recognition system to extract information from high-resource monolingual datasets. \cite{meta2} applies meta learning approach for low-resource ASR, in order to learn better initialized parameters from many other languages for the unseen target language. \cite{meta3} introduces the meta-learning paradigm to a few-shot spoken intent classiﬁcation task and achieves comparable performance to data-rich models.

In this work we utilize the MAXL algorithm to train ASR and NLU jointly. In order to make it workable, we study different interfaces from NLU network to ASR-SLU network to make the entire network differentiable. The differences between our work and the original MAXL are: 1) MAXL only focuses on the primary task and the training of auxiliary task aims at enhancing the performance of primary. However, we are more concerned about the performance of the upstream outputs on downstream tasks. Therefore we consider the evaluation on the auxiliary task and insist on seeking for a differentiable interface to update the auxiliary network. The interface in the raw MAXL is relatively simple, while it is much sophisticated in our task, which includes either classification and sequence labelling task and has variable lengths in the output layer. 
2) The inputs of the two networks in MAXL are the same, while in our model we use different inputs to take advantage of the manual transcripts as a teacher predictor thus to improve the SLU prediction. 3) It is recommend in the work of MAXL that the auxiliary task should be more complicated than the primary task to gain benefits, while in our implementation, the search space of ASR is much larger than NLU and it still works.

\section{Method}

\begin{figure}[t]
 \centering
 \includegraphics[width=8cm]{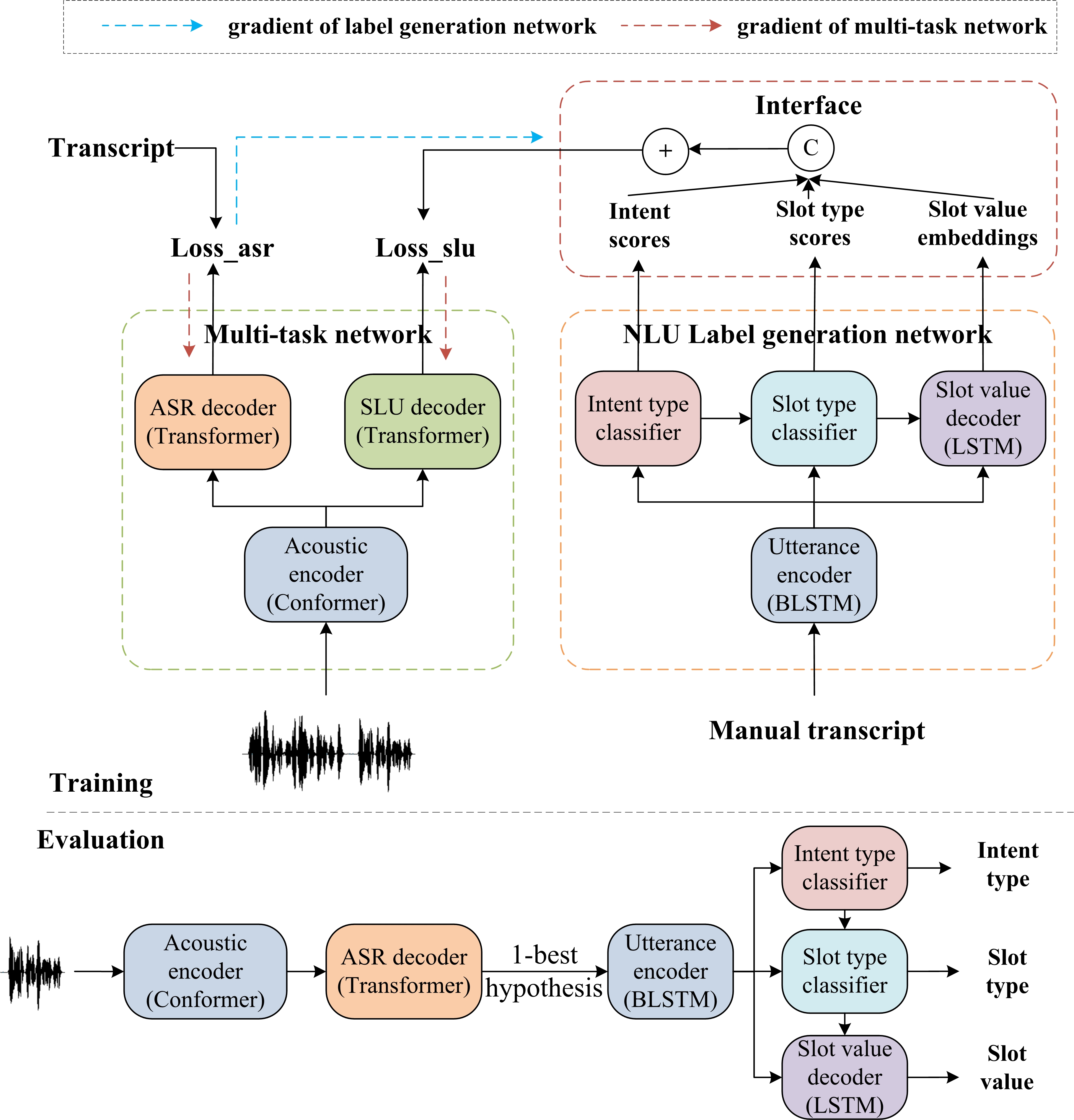}
 \caption{The proposed MAXL based SLU model.}
 \label{fig1}
\end{figure}

\subsection{Model}
The proposed approach consists of two networks: 1) a multi-task network to train ASR task and SLU task synchronously, and 2) a NLU network to predict semantic tags for the SLU task. The input of the multi-task network is speech signal while the input of NLU network is manual transcript. The output of NLU network is delivered to the output layer of SLU task as a target to make the SLU prediction close to the NLU prediction which is relatively data-rich. An interface is inserted between the two networks to make the entire model differentiable, which concatenates the possibilities predicted by NLU and adds them up for a fixed-length. During the evaluation stage, we send the ASR 1-best output to the NLU model, since the SLU decoder fails to gain a readable result due to the summation operation in the interface. The outputs of ASR are more suitable for the NLU task. Meanwhile, the NLU model is also enhanced though semantic tags are not adopted in the MAXL training stage.

\subsubsection{Multi-task Network}
The multi-task network comprises one shared encoder and two decoders for ASR task and SLU task separately. The two decoders have different vocabulary sizes since they aim at different target space. It is constructed based on an end-to-end ASR model with conformer blocks as encoder and transformer blocks as decoder. An additional transformer decoder is appended for SLU task. The ASR related modules are initialized via a large-scale pre-trained model then fine-tuned via the low-resource SLU data together with SLU module.

\subsubsection{Label Generation network}
The label generation part is a traditional NLU model with a BLSTM encoder and a LSTM decoder \cite{hierachical}. The intent classifier, the slot type classifier and the slot value decoder share an utterance encoder. They predict intent, slot and value one by one. Specifically, the prediction of the former will be sent to the latter as features with the utterance. We select this architecture since it achieves state-of-the-art performance on the open-source Chinese SLU dataset that we use. 

\subsubsection{Interface}
The output of the label generation network is a list of intent-slot-value triples, the length of which is variable since the number of slots is utterance dependent. However, the length of the decoder output in the multi-task network should be equal inside a mini-batch. Besides, the predicted label should support gradient flow for the updating of label generation network with respect to the ASR loss. Therefore, the argmax operation on the softmax output is not allowed when it is delivered to the multi-task network since it will break the gradient flow. We consider several interfaces from the two aspects: fixed length and gradient flow, as listed in Table~\ref{tab1}. 

\begin{table}[ht]
\centering
\caption{The interfaces between the two networks and their attributes.}
  \label{tab1}
\begin{tabular}[H]{lcc}
\hline
             & Fixed & Gradient   \\
             & length & flow  \\
\hline
List  & $\times$  & $\times$   \\
Sequence  & \checkmark  & $\times$   \\
NER tag          &  \checkmark  &  $\times$   \\
Softmax     & $\times$    & \checkmark   \\
Sum of softmax     & \checkmark   & \checkmark     \\
Append intent and slot types   & \checkmark  & \checkmark  \\
\hline
\end{tabular}
\end{table}

\begin{itemize}
\item \textbf{List}: ``[[intent-slot-value]$_1$,[intent-slot-value]$_2$,...]'', this is the final format of the NLU prediction that stores the predicted triples in a list, which has variable-length and the argmax or top-one operation breaks the gradient flow during the NLU network updating. 
\item \textbf{Sequence}: ``value$_1$ value$_2$ ...'', this is a concatenation of the slot values, which is able to meet fixed length via padding, but does not support gradient flow either.
\item \textbf{NER tag}: ``[1 1 0 1 1 1]'', it marks the slot tokens as 1 and others as 0 in an utterance, which has the same length with the utterance. Similarly the gradient flow is also broken by top-one operation.
\item \textbf{Softmax}: $\mathbb{R}\in N\times L\times V$, $N$ is the slot number, $L$ is the utterance length and $V$ is the vocabulary size. This refers the output of softmax before beam search. It maps the tokens in the slots to vocab size embeddings which can be adopted as word embeddings. However, the length of these embeddings is still determined by slot number which is not fixed. 
\item \textbf{Sum of softmax}: $\mathbb{R}\in L\times V$. We employ a sum operation on the softmax embeddings along the first dimension and gain a fixed-length feature which is also differentiable. 
\item \textbf{Append intent and slot types}: $\mathbb{R}\in L\times (V+I+S)$, $I$ is the number of intent types and $S$ is the number of slot types. We expand the slot value feature with intent scores and slot type scores per token to supply a contextual reference.
\end{itemize}

\subsection{Training}
The training procedure per epoch is divided into two stages. The first stage is an ordinary multi-task learning problem, where the targets of ASR are manual transcripts and the targets of SLU are the predicted semantic tags from a NLU network. This stage can be seen as the auxiliary-training step in meta learning which encourages ASR to focus on the semantic related part. The loss function is defined as:
\begin{equation}
  \mathcal{L}_{\theta_{1}}=\mathcal{L}(f_{\theta_{1}}^{asr}(x_{(i)},y_{(i)}^{asr})+\mathcal{L}(f_{\theta_{1}}^{slu}(x_{(i)},g_{\theta_{2}}^{nlu}(y_{(i)}^{asr})))
\end{equation}
in which $i$ denotes the batch index, $x$ represents the acoustic input feature, $y^{asr}$ is the manual transcript, $\theta_{1}$ refers to the parameters of the multi-task network $f$ and $\theta_{2}$ means the parameters of the label generation network $g$.

In the second stage the NLU network is updated by computing its gradients with respect to the ASR loss in the multi-task network. Second derivatives are generated due to a previous update of the multi-task network, thus this stage can be seen as the meta-training step in meta learning. Firstorder \cite{maxl} \cite{darts} is an approximate implementation of MAXL algorithm, which is based on the finite difference method and successfully speeds up 4 - 6 times training time. We 
adopt this second derivative trick and compare it with standard implementation in MAXL. The loss function is as follows:
\begin{equation}
 \mathcal{L}_{\theta_{2}}=\mathcal{L}(f_{\theta_{1}^{+}}^{asr}(x_{(i)},y_{(i)}^{asr})+\mathcal{H}(\bar{y}_{(i)}^{nlu})
\end{equation}
\begin{equation}
\begin{split}
& \theta_{1}^{+}=\theta_{1}- \\
& \alpha\nabla_{\theta_{1}}(\mathcal{L}(f_{\theta_{1}}^{asr}(x_{(i)},y_{(i)}^{asr}))+\mathcal{L}(f_{\theta_{1}}^{slu}(x_{(i)},g_{\theta_{2}}^{nlu}(y_{(i)}^{asr})))) \\
\end{split}
\end{equation}
Notably, the computational graph of the updated multi-task network $\theta_{1}^{+}$ should be retained to compute the derivatives with respect to the NLU network. An entropy loss $\mathcal{H}(\bar{g}_{\theta_{2}}^{nlu})$ is appended to prevent collapsing and to encourage the network to utilize all NLU classes, which is calculated based on the average prediction in a mini-batch of the NLU network.
\begin{equation}
 \mathcal{H}(\bar{g}_{\theta_{2}}^{nlu})=\sum_{k=1}^{k}\bar{g}_{\theta_{2}}^{k}\log\bar{g}_{\theta_{2}}^{k}
\end{equation}
\begin{equation}
\bar{g}_{\theta_{2}}^{k}=\frac{1}{N} \sum_{n=1}^{N}g_{\theta_{2}}^{k}[n]
\end{equation}
 $K$ is the dimension of the NLU prediction and $N$ is the batch size. 

\section{Experiments}

In this section, the proposed method is first compared with other training method for SLU. Then the effects of some key components such as the second derivative trick, different interfaces and the pre-trained model are investigated. Finally, the method is implemented in a semi-supervised manner to see the potency of the proposal to decrease the requirement of semantically annotated data.

\subsection{Dataset and Experimental Setup} 
The experiments are carried out on an open-source Chinese SLU dataset CATSLU\cite{catslu}, which is used in the 1st Chinese Audio-Textual Spoken Language Understanding Challenge. It contains speech recordings and text transcripts, annotated with semantic intent-slot-value labels. The CATSLU-MAP dialogues in the map navigation domain are selected in this work, with a total of 5093 utterances in training set, 1578 utterances in test set and 921 utterances in development set. The multi-task network consists of 12 conformer blocks as the encoder followed by 6 transformer blocks as each decoder for ASR and SLU. The feed-forward layer contains 2,048 units and the number of attention heads is 4, with 256 units in each head. The hidden vector dimension of the LSTM network used in NLU is 256 and the word embedding dimension is 200.

\subsection{Evaluation Metrics} 
The output of SLU module is a sum of the semantic predictions which is not able to revert to the list of intent-slot-value triples. Therefore the proposal is evaluated by the performance of the NLU network predicted from the ASR 1-best hypothesis. The Character Error Rate(CER) of the ASR model is also reported as an auxiliary metric. Meanwhile, the performance of the NLU model based on manual transcript is also presented to demonstrate whether the unsupervised label generation network is able to extract knowledge from the supervised task.

\subsection{Results}
\subsubsection{Method Comparison}
In this test, the NLU model is pre-trained through all the labeled CATSLU-MAP data, and the ASR model is pre-trained by 6000 hours of mandarin speech. During the MAXL training, only the manual transcripts and speech for ASR are employed. The results are presented in Table~\ref{tab2}. From the first row we can see our basic ASR model performs poorly on the NLU task. The end-to-end method (taking the outputs of Gumble softmax as the interface) obtains slight improvements on both CER and F1-score. As the other baseline, we fine-tune the ASR model via the speech in CATSLU-MAP and deliver the outputs to NLU model. The CER declines significantly and the F1-score of NLU increases by 4.33\%. We utilize either standard MAXL algorithm or first-order algorithm to calculate second derivative, and both of them surpasses the fine-tuned results in both CER and F1-score. First-order algorithm is selected since it performs better.
To further verify the effect of NLU results on ASR model, we train a multi-task network without the label generation network. The ground truth semantic tags are used as the targets of SLU. As we can see in Table~\ref{tab2}, the proposal(First-order) achieves comparable results with the multi-task network driven by ground truth SLU annotation. This verifies the performance of the NLU predictor and the effectiveness of it on the ASR prediction. And more importantly, the proposal is more extensible since the training does not require any further semantic labels.

\begin{table}[ht]
\centering
\caption{The performance comparison between the proposal and baselines.}
  \label{tab2}
\begin{tabular}{lccc}
\hline
             & CER & F1-score \\ 
Baseline     &   28.76         &  52.00 \\ 
End-to-end  &    27.75      &  53.12 \\ 
Fine-tuned   &   21.45         & 56.33 \\ 
Proposal(MAXL)       &    21.21    &   58.27 \\
Proposal(First-order)     &    \textbf{21.15}       &  \textbf{59.64}   \\
Multi-task   &    21.30        &  59.25    \\
\hline
\end{tabular}
\end{table}

Although the decrease on CER is insignificant, the performance of the ASR output on the NLU task gains a more obvious improvement, which implies that the ASR is more suitable for the downstream task. To further confirm that, we decode the training set via the updated ASR model and trained a new NLU model based on the hypotheses. The results in Table~\ref{tab3} show that the performance of the retrained model by ASR hypotheses is close to the one trained by manual transcripts on the current test set.

\begin{table}[ht]
\centering
\caption{The CER of training set and the performance of NLU model trained by the ASR hypotheses.}
  \label{tab3}
\begin{tabular}{lcc}
\hline
             & CER & F1-score   \\
Fine-tuned     &  10.57        &  53.30      \\         
Proposal(First-order)  &   \textbf{9.14}       &  \textbf{56.32}  \\
\hline
\end{tabular}
\end{table}

\subsubsection{Interface}
The interfaces with variable-length are unavailable in the proposed method. Therefore we compare the remaining four interfaces with fixed-length. The two non-differentiable interfaces are still implemented to test the effects of differentiability. As shown in Table~\ref{tab4}, the performances of the four interfaces are close. However, differentiable interfaces is able to achieve better NLU models. Although the improvement is not very obvious, it still verifies that the NLU network is able to extract knowledge from the multi-task network. Interface 4 proves that the appending of intent and slot scores is helpful for the NLU prediction.

\begin{table}[ht]
\centering
\caption{The performance of the proposal with different interfaces. Inter1, Inter2, Inter3 and Inter4 refer to the Interfaces of Sequence, NER tag, Sum of softmax and Append intent and slot types, respectively.}
  \label{tab4}
\begin{tabular}{lccc}
\hline
             & CER & F1-score   & F1-score   \\  
             &   & (asr 1-best)   &  (transcript) \\
Inter1  &   21.22         &  59.01   &  91.22 \\
Inter2  &    21.30         &    58.95  & 91.22 \\    
Inter3     &   \textbf{21.12}       &  59.00   &  \textbf{91.34}   \\         
Inter4   &     21.15     &  \textbf{59.64}    &  91.28 \\  
\hline
\end{tabular}
\end{table}

\subsubsection{Pre-trained model}
Furthermore, we investigate whether the pre-trained model is crucial for both ASR and NLU task. The results without pre-trained models are listed in Table~\ref{tab5}, which demonstrates that both the ASR and NLU pre-trained models are crucial for the MAXL based SLU system, especially the NLU pre-trained model. Larger language models such as Bert will be introduced to support a better initialization in the future when we adopt a transformer based encoder for NLU.

\begin{table}[ht]
\centering
\caption{The results without a ASR pre-trained model or without a NLU pre-trained model.}
  \label{tab5}
\begin{tabular}{lcc}
\hline
             & CER  & F1-score  \\
w/o ASR pre-trained     &    89.87   &  24.33       \\       
w/o NLU pre-trained   &   87.42     & 0.04     \\  
\hline
\end{tabular}
\end{table}

\subsubsection{Semi-supervised}
At last, we implement the proposal in a semi-supervised way that only half semantic tagged data participates in the pre-training of NLU model, and the remaining half is adopted in the MAXL training without semantic tags. The results in Table~\ref{tab6} prove that the proposal is effective when there are no semantic annotations for the incremental training. The limitation is that the appended data should come from the same domain with the NLU training data, which means that the prediction of NLU on the current utterance should be reliable enough otherwise it will supply a bad guidance. 

\begin{table}[ht]
\centering
\caption{The results that only half semantic tags are adopted.}
  \label{tab6}
\begin{tabular}{lcc}
\hline
       & CER & F1-score   \\
 half data pre-train       &23.45  &    54.98   \\     
     Proposal(First-order)  & \textbf{23.08}         &   \textbf{55.93}    \\ 
\hline
\end{tabular}
\end{table}

\section {Conclusion}
This work proposed a MAXL based SLU model which decrease the requirement of semantic tags on speech data. The joint training between ASU and NLU enhances the quality of ASR outputs for NLU task and extracts knowledge from the supervised network for the low-resource task. The results are consistent with our hypothesis that the interaction between ASR and NLU is not unidirectional that the ASR output effects the NLU performance, meanwhile the NLU prediction should feedback to guide the emphasis of ASR. 
In the future, the proposal will be tested on more open-source datasets with more complex network structures.

\bibliographystyle{IEEEtran}

\bibliography{mybib}


\end{document}